\documentclass{aastex}

\shorttitle{Pulsed Jets }
\shortauthors{Cerqueira \& de Gouveia Dal Pino }

\begin{document}

\title{On the Influence of Magnetic Fields on the Structure of Protostellar  
Jets}

\author{A.H. Cerqueira \& E.M. de Gouveia Dal Pino  }
\affil{Instituto Astron\^{o}mico e Geof\'{\i}sico, Universidade de
S\~{a}o Paulo, \\ Av. Miguel St\'efano 4200, (04301-904) S\~{a}o Paulo -
SP, BRAZIL \\ adriano@iagusp.usp.br, dalpino@iagusp.usp.br}

\begin{abstract}

We here present the first results of fully three-dimensional (3-D)
MHD simulations of radiative cooling pulsed (time-variable) jets for a
set of parameters which are suitable for protostellar outflows. Considering
different initial magnetic field topologies in approximate $equipartition$
with the thermal gas, i.e., (i) a longitudinal, and (ii) a helical
field, both of which permeating the jet and the ambient medium; and
(iii) a purely toroidal field permeating only the jet, we find that the
overall morphology of the pulsed jet is not very much affected by the
presence of the different magnetic field geometries in comparison to a
nonmagnetic calculation.  Instead, the magnetic fields tend to affect
essentially the detailed structure and emission properties behind the shocks at
the head and at the pulse-induced internal knots, particularly for the
helical and toroidal geometries. In these cases, we find, for example,
that the $H_\alpha$ emissivity behind the internal knots can be about
three to four times larger than that of the purely hydrodynamical jet.
We also find that some features, like the nose cones that often develop
at the jet head in 2-D calculations involving toroidal magnetic fields,
are smoothed out or absent in the 3-D calculations.

\end{abstract}

\keywords{ISM: jets and outflows -- ISM: Herbig-Haro objects -- MHD --
star: formation}

\section{Introduction}

Low-mass young stellar objects are known to produce collimated optical
outflows (the  Herbig-Haro jets; hereafter, HH jets) that  may extend from
a few 1000 AU (e.g., Bally, O'Dell \& McCaughrean 2000), to very large
parsec-length scales (Heathcote, Reipurth \& Raga 1998; Bally \& Devine
1997), and propagate with v $\sim$ few 100 km/s into the ambient medium.
Immersed in these jets, there is a chain of bright emission knots whose
bow shock morphology and  high spatial velocity (e.g., Reipurth 1989;
Zinnecker, McCaughrean \& Rayner 1998) all indicate that they are shocks
that arise from the steepening of velocity (and/or density) fluctuations
in time in the underlying, supersonic outflow. Strong support for this
conjecture has been given by theoretical studies which have confirmed that
traveling shocks created in this manner reproduce the essential properties
of the observed knots (e.g., Raga {\it et al.} 1990; Kofman \& Raga 1992;
Stone \& Norman 1993; de Gouveia Dal Pino \& Benz 1994; V\"olker {\it
et al.} 1999; Raga \& Cant\'o 1998, de Gouveia Dal Pino 2001).

There is increasing evidence that magnetic fields play a major role
both  in the production and collimation of  these HH jets and the most
promising mechanism for  their launching involves magneto-centrifugal
forces associated either with the accretion disk that surrounds the
star (e.g. K\"onigl \& Pudritz 2000; Kudoh, Matsumoto \& Shibata 1998),
or with the disk-star boundary (in the X-winds; e.g. Shu {\it et al.}
1994). Direct observations of magnetic fields ($\bf{B}$) are difficult
and only recently Ray {\it et al.} (1997) have obtained the first direct
evidence, based on polarization measurements, of $\bf{B}$-fields $\sim$
1 G in the outflow of T Tau S at a distance of few tens of AU from the
source.  Using magnetic flux conservation and typical jet parameters,
this latter result would imply a plasma $\beta \, = \, p_{j,gas}/(B^2/8
\pi) \, \simeq \, 10^{-3}$ for a toroidal field configuration, and $
\sim 10^{3}$, for a longitudinal field, at distances $\sim 0.1$ pc.
Since  the outflowing material does not seem to be highly resistive
(e.g., Frank {\it et al.} 1999),  these figures indicate that magnetic
fields may also play a relevant role on the outer scales of the flow.

Lately,  motivated  by these pieces of information, several MHD studies
of overdense, radiative cooling jets have been developed in an attempt
to find out possible  signatures of $\bf{B}$-fields on the large scales
of the HH outflows.  In previous investigations (de Gouveia Dal Pino
\& Cerqueira 1996; Cerqueira, de Gouveia Dal Pino \& Herant 1997,
hereafter Paper I; Cerqueira \& de Gouveia Dal Pino 1999, hereafter
Paper II), we have employed a three-dimensional (3-D) {\it Smoothed Particle
Magnetohydrodynamic} (hereafter SPMHD) code to investigate steady-state
cooling jets. More recently, two-dimensional, axis-symmetric calculations
of magnetized steady and pulsed jets have been developed by Frank {\it
et al.} (1998, 1999), Gardiner {\it et al.} (2000); Stone \& Hardee
(2000; hereafter SH); O'Sullivan \& Ray (2000; hereafter OR), Lery \&
Frank (2000), and Gardiner \& Frank (2000; hereafter, GF).  In this
letter, we try to extend these prior studies by presenting results of
fully 3-D calculations of MHD radiatively cooling,
$pulsed$ jets, considering different magnetic field topologies.  In a
forthcoming paper (Cerqueira \& de Gouveia Dal Pino 2001, hereafter CG01),
we explore the emission properties and structure of pulsed jets in more
detail and cover a more extensive range of parameters.

\section{Numerical Method, Input Conditions, and Results}

To simulate the effects of magnetic fields on pulsed jets, we solve
the ideal MHD equations using our 3-D SPMHD code previously developed
for the investigation of purely hydrodynamical (e.g., de Gouveia Dal
Pino \& Benz 1993, hereafter GB93, 1994; de Gouveia Dal Pino 1999,
2001), and also magnetized steady-state jets (e.g., Paper I and II).
(See Paper II and CG01 for a detailed description of the code and method.)

Our computational domain is a rectangular box with dimensions $-30 R_j
\le$ x $\le 30 R_j$, and $-10R_j \le$ y,z $\le 10 R_j$, where $R_j$
is the initial jet radius (which is also the code distance unit). The
Cartesian coordinate system has its origin at the  center of the
box and the jet is continuously injected into the bottom of the box
[at ${\bf r}=(-30R_j,0,0)$]. Inside the box, the SPH particles are
initially distributed on a cubic lattice. An outflow boundary condition
is assumed for the boundaries of the box.  The particles are smoothed out
by a spherically symmetric kernel function of width $h$, and the initial
values of $h$ were chosen to be $0.4 R_j$ and $0.2R_j$ for the ambient and
jet particles, respectively, so that we have up to 400,000 SPH particles
at the beginning of the calculation. The adiabatic index of the ambient
medium and the jet is assumed to be $\gamma=5/3$, and an ideal equation
of state is used. The radiative cooling, due to collisional excitation
and recombination, is implicitly calculated using a time-independent
cooling function for a gas of cosmic abundances cooling from $T \simeq
10^6$ to $10^4$ K (see, e.g., de GB93).  We adopt a sinusoidal velocity
profile for the pulsing jet at the inlet: $v_0(t)= v_j [1 + A {\rm
sen}({{2 \pi}\over P}  t)]$, where $v_j$ is the mean jet speed, $A$
is the velocity amplitude and $P$  the period of the oscillation.

Given the present uncertainties related to the real orientation and
strength of the magnetic field in HH jets, we assume, as in
previous work, three different initial magnetic field configurations.
In one case, we consider an initially  constant longitudinal magnetic
field parallel to the jet axis permeating both the jet and the ambient
medium ${\bf{B}} =(B_0,0,0)$ (e.g., Paper I).  The second geometry
adopted here is a force-free helical magnetic field which also extends
to the ambient medium and whose radial functional dependence is given
in Fig. 1 of Paper II, and by equations 19 to 21 of Todo {\it et al.}
(1993). In this case, the maximum strength of the magnetic field in
the system corresponds to the magnitude of the longitudinal component
of the field at the jet axis ($B_0 \simeq 83\mu$G ). The azimuthal
component attains a maximum value ($B_\phi = 0.39 B_0$) at $\sim 3 R_j$
(see Paper II). The third geometry adopted is a purely toroidal magnetic
field permeating the jet only, with a radial functional form given by
equation 5 and Fig. 1 of SH (see also CG01).  The field in this case
is zero on the jet axis, achieves a maximum strength ($\approx 170\mu$G)
at a radial position in the jet interior ($R_m \approx 0.9 R_j$), and
returns to zero  at the jet surface. This field corresponds to a uniform
current density inside the jet and a return current at the surface of
the jet. In the first two magnetic field configurations, the jet is
assumed to have an initially constant gas pressure ($p_j$) which is in
equilibrium with the pressure in the ambient medium ($\kappa = p_j/p_a$
= 1). In the toroidal configuration, in order to ensure an initial
magnetostatic equilibrium, the jet gas pressure has a radial profile
with a maximum at the jet axis $p_j(0)=2.07$ $p_j(R_j)$, and $p_j(R_j)=
p_a$ at the jet surface (see Fig. 1 and equations 7 and 8 of SH).

The adopted parameters for the simulations are appropriate to conditions
found in HH jets. We take a number density ratio between the
jet and the ambient medium $\eta=n_j/n_a=5$, $n_j=1000$ cm$^{-3}$,
an average ambient Mach number $M_a = v_j/c_a=15$ (where $v_j$ is
the average jet velocity and $c_a \simeq 16.6 $ km s$^{-1}$ is the
ambient sound speed), $v_j \simeq $ 250 km s$^{- 1}$, and $R_j = 2
\times 10^{15}$ cm. The assumed amplitude and period of the velocity
oscillations are, $A=0.25 v_j$, and $P=0.54 t_d$ (where $t_d=R_j/c_a
\approx 38$ years corresponds to the transverse jet dynamical time),
respectively.  In the MHD calculations, we assume an initial maximum
$\beta_j=p_{j}/p_{B,j} \simeq 1$.  The corresponding initial average,
magnetosonic jet Mach number is $M_{ms,j} = v_j/(v_{A,j}^2 + c_j^2)
\simeq  25$. 

Figure 1 displays the midplane density contours (left) and the velocity
field distribution (right), for four supermagnetosonic, radiatively
cooling, $pulsed$ jets in their early evolution, after they have
propagated over a distance $\approx 22 R_j$, at $t/t_d = 1.75$. The
top jet is purely hydrodynamical (HD) ($\beta_j = \infty$, $M_{ms,j} =
34$); the second jet has an initial constant longitudinal magnetic field
configuration; the third jet has an initial helical magnetic field, and
the bottom jet has an initial toroidal field. In the HD case, the ratio of
the radiative cooling length in the post-shock gas behind the bow shock to
the jet radius is $q_{bs}=d_{cool,bs}/R_j \simeq 1.7$ and the correspondig
ratio behind the jet shock is $q_{js} \simeq \eta^{-3}q_{bs} \simeq 0.01$
(e.g., GB93). These cooling length parameters imply that within the head
of the jet, both the ambient shocked gas and the shocked jet material
are subject to rapid radiative cooling. Similar initial conditions are
obtained for the MHD cases.

At the time depicted, the leading {\it working} {\it surface} at the
jet head is followed by 2 other features, and a third pulse just entering
the system. Like the leading working
surface, each internal feature consists of a double-shock structure,
an upstream reverse shock  that decelerates the high velocity material
entering the pulse, and a downstream forward shock  sweeping up the low
velocity material ahead of the  pulse.  Each of these internal working
surfaces (IWS) or knots, widens and broadens as it propagates downstream
and squeezes shocked material sideways into the cocoon.  At this time,
no IWS has reached the jet head yet. Later on, however, the
IWSs, one by one, reach  the terminal working surface and collide with
it. They are thus disrupted by the impact and their debris are partially
deposited into the cocoon thus providing a complex fragmented structure
at the head as required by the observations (see CG01).

Fig. 1 indicates that the overall morphology and dynamics of the
3-D pulsed jet is not very much affected by the presence of the
different magnetic field configurations when compared to the baseline
HD calculation. Instead, the distinct magnetic field profiles  tend to
alter only the  detailed structure behind the shocks at the head and
internal knots, particularly in  the helical and purely toroidal cases
\footnote{We note that this condition persists also in the later evolution
of the jets; see CG01.}. For example, in the pure HD jet (Fig. 1, top)
we can identify, as in previous work (e.g., GB93; Paper II), the formation
of  a cold dense shell  at the head (with maximum density $n_{sh}/n_{a}
\sim 119$), due to the cooling of the shock-heated material. It becomes
Rayleigh-Taylor (R-T) unstable (GB93) and breaks into clumps which can be
also  identified in the shell of the MHD jet with initial longitudinal
field, but not in the helical or toroidal cases.  In these later
cases, the toroidal component of the magnetic field which is initially
less intense than the longitudinal component is strongly amplified by
compression in the shocks at the head (by a factor $\gtrsim 2$). This
amplification reduces the density enhancement and increases the cooling
length behind the jet shock. As a consequence, the shell tends to be
stabilized against the R-T instability and the fragmentation and formation
of clumps is thus inhibited (as seen in previous studies of steady jets;
Paper I and II).

Figure 1 also indicates that, at least for the initial conditions examined
here, no {\it nose} {\it cones} develop at the head of the magnetized jets, not
even in the case with initial purely toroidal field (see also CG01). Such
features are extended narrow plugs that often develop in 2-D calculations
(e.g., OR; SH), at the head of magnetized jets with  toroidal fields. They
result from the accumulation of the  shocked gas between the jet shock
and the bow shock which is  prevented from escaping sideways into the
cocoon by hoop stresses associated with the toroidal field. The absence
of these nose cones in the 3-D calculations could be explained by the
fact that, as the magnetic forces are intrinsically three-dimensional,
they cause part of  the material to deflect in a third direction,
therefore, smoothing out the strong focusing of the shocked material
that is otherwise detected in the 2-D calculations.

Figure 2 depicts several physical quantities along the beam axis for the
jet of Fig. 1 with initial helical magnetic field, but at a later time
($t/t_d \simeq 3$) when then two more pulses have entered the system
and developed new IWSs, and the first IWS has already merged  with
the jet head.  We see (top panel) that the initial sinusoidal velocity
profile impressed on the flow at the inlet steepens  into the familiar
sawtooth pattern seen in earlier studies of oscillating jets (e.g.,
Raga \& Kofman 1992) as the faster material catches up with the slower,
upstream gas in each pulse.  The downward sharp density spikes within
each IWS (second panel of Fig. 2) first grow and then fall off in the
downstream direction. This effect has been documented in the literature
and is caused by the progressive expelling of material sideways into
the cocoon (see, e.g., de Gouveia Dal Pino 2001).  We also find that
this decay is more pronounced in 3-D jets than in 2-D (see CG01).
The third panel shows that the toroidal component of the magnetic field
($B_{\phi}$) sharpens within the knots and rarefies between them, while
the longitudinal component ($B_{//}$) (forth panel) is stronger between
the knots. (We note that the toroidal magnetic field component in
the IWSs is amplified by a factor of $\approx$ 1.5 for both the helical
and toroidal configurations.)

The results above suggest that although the distinct $\bf{B}$-geometries
do not significantly affect the global characteristics of the flow, they
can modify the details of the emission structure, particularly for the
helical and toroidal cases. As an example, Figure 3 shows the $H_{\alpha}$
intensity along the jet axis evaluated within the IWSs for the different
magnetized jets of Fig. 1 which are compared to the $H_{\alpha}$ intensity
of the baseline hydrodynamical jet.  The $H_{\alpha}$ intensity is known
to be strong behind the knots of the observed HH jets and
is predicted to have a strong dependence with the shock speed $v_s$,
$ I_{H{\alpha}} \propto \rho_{d} v_s^{3.8} $, where $\rho_{d}$ is the
downstream pre-shock density (e.g., Raga \& Cant\'o 1998). We have used
this relation and the results of the simulated jets of Fig. 1 to evaluate
the intensity ratios of Fig. 3.  We note, as expected, that the intensity
ratio does not differ too much from unity in the case of the magnetized
jet with initial longitudinal field, while for the helical and toroidal
field cases, the $H_{\alpha}$ intensity can be about three to four times
larger than that of the HD case.

\section{Discussion and Conclusions}

Consistently with previous work (e.g., Paper I and II; OR; SH;
Gardiner {\it et al.} 2000), the present  three-dimensional (3-D)
MHD simulations of radiatively cooling, pulsed  jets have shown that
the effects of magnetic fields are dependent on the field-geometry
(which, unfortunately, is still poorly determined from observations).
For example, the  presence of a helical or a toroidal field tends to
affect more the characteristics of the fluid, compared to the purely HD
calculation, than a longitudinal field.  However, the relative differences
previously detected in 2-D simulations  involving distinct magnetic
field geometries (OR; SH), seem to decrease in the 3-D calculations
(see also CG01).  In particular, we have found that features, like
the  nose cones, that often develop at the jet head in 2-D calculations
involving toroidal magnetic fields, are absent in our 3-D calculations.
Consistently,  SH have also recently argued that nose cones should be
unstable in 3-D. This result is also consistent with observations which
show no direct evidence for nose cones at the head of HH jets.

In 3-D calculations, magnetic fields which are initially in near
equipartition with the gas, tend to affect essentially  the detailed
structure behind the shocks at the head and internal knots, mainly
for the helical and toroidal topologies. In such cases, the $H_\alpha$
emissivity behind the internal knots can increase by a factor up to four
relative to that in  the purely hydrodynamical jet.  Our results also show
that in the presence of a helical magnetic field, the toroidal component
tends to be dominant within the internal knots while the longitudinal
component tends to be larger between them.  This result is in agreement
with recent 2-D calculations (OR; GF) and could in principle, be checked
observationally as a diagnostic of helical geometries. Helical fields
are specially attractive for HH jets as they are the predicted geometry
from magneto-centrifugally launched models (GF).

Finally, the results above suggest that further observations with the
construction of polarization maps are urgently required for a real
comprehension of the magnetic field structure of HH jets,  as
the sensitivity to its topology (and strength) seems to be  relevant.
Further 3-D MHD studies are also required since the detailed structure
and emission properties of the jets are affected by multidimensional
effects when magnetic forces are present (see also Paper I and CG01).

\acknowledgments

We thank the anonymous referee for many useful suggestions, which
have improved the final version of the paper.
A.H.C. and E.M.G.D.P. acknowledge the Brazilian agencies FAPESP and
CNPq for support. The simulations were performed on a cluster
of Linux-based PC's, as well in a DEC 3000/600 AXP workstation, whose
purchases were made possible by FAPESP.

\newpage

\figcaption[F1.eps]{ Midplane density contours (left) and distribution of
velocity vectors (right) for (from top to bottom): a purely hydrodynamical
pulsed jet; an MHD jet with initial longitudinal magnetic field; an MHD
jet with initial helical magnetic field; and an MHD jet with initial
toroidal magnetic field, at a time $t/t_d=1.75$ ($t_d=R_j/c_a \simeq 38$
years). The initial conditions are: average $M_a=15$, $v_0 =v_j[1+A{\rm
sin}(2\pi t / P)]$, with $v_j \simeq 250$ km s$^{-1}$, $A=0.25 v_j$
and $P=0.54 t_d$, $\eta= n_j/n_a =5$, $\beta =  8\pi p/B^2 = \infty$
for the HD model, and maximum $\beta \simeq 1$ for the MHD models. The
maximum density in each model (from top to bottom) is: 119, 115, 124
and 175 (1 c.u. $= 200$ cm$^{-3}$). The $x$ and $z$ coordinates are in units
of $R_j$. The jet is injected into the computational domain at
$x = -30R_j$. \label{fig1}}

\figcaption[F2.eps]{From top to bottom: velocity, density, toroidal
magnetic field component, and longitudinal magnetic field profiles
along the beam axis for the jet of Fig. 1 with initial helical magnetic
field (third jet from top of Fig. 1).  Time displayed $t/t_d \simeq 3 $.
Note that the coordinate along the jet axis ($d$) has been shifted from $
-30 R_j$ to $0$.  (The noise we see near the head results from the impact
of the first internal knot with the head.)  \label{fig2}}

\figcaption[F3.eps]{ The ratio between the $H_{\alpha}$ intensity along
the jet axis within the IWSs for the different magnetized jets of Fig. 1
and the the $H_{\alpha}$ intensity of the purely hydrodynamical jet: MHD
model with longitudinal field (MHD L, squares), MHD model with helical field
(MHD H, bullets) and MHD model with toroidal field (MHD T, stars).
The $d$ coordinate is the same as in Fig. 2.
\label{fig3}}

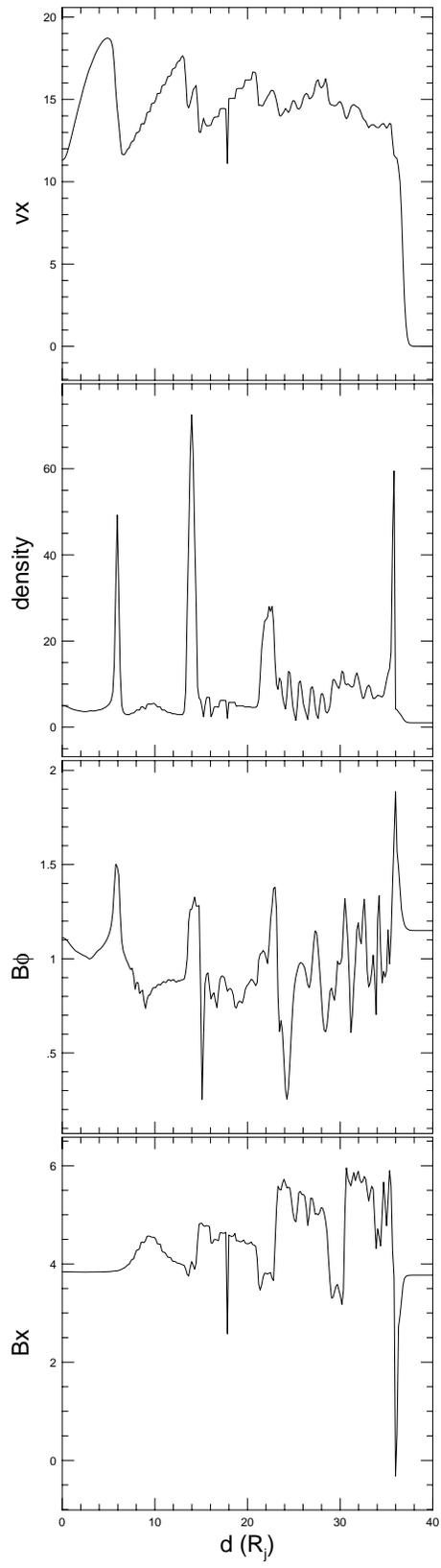
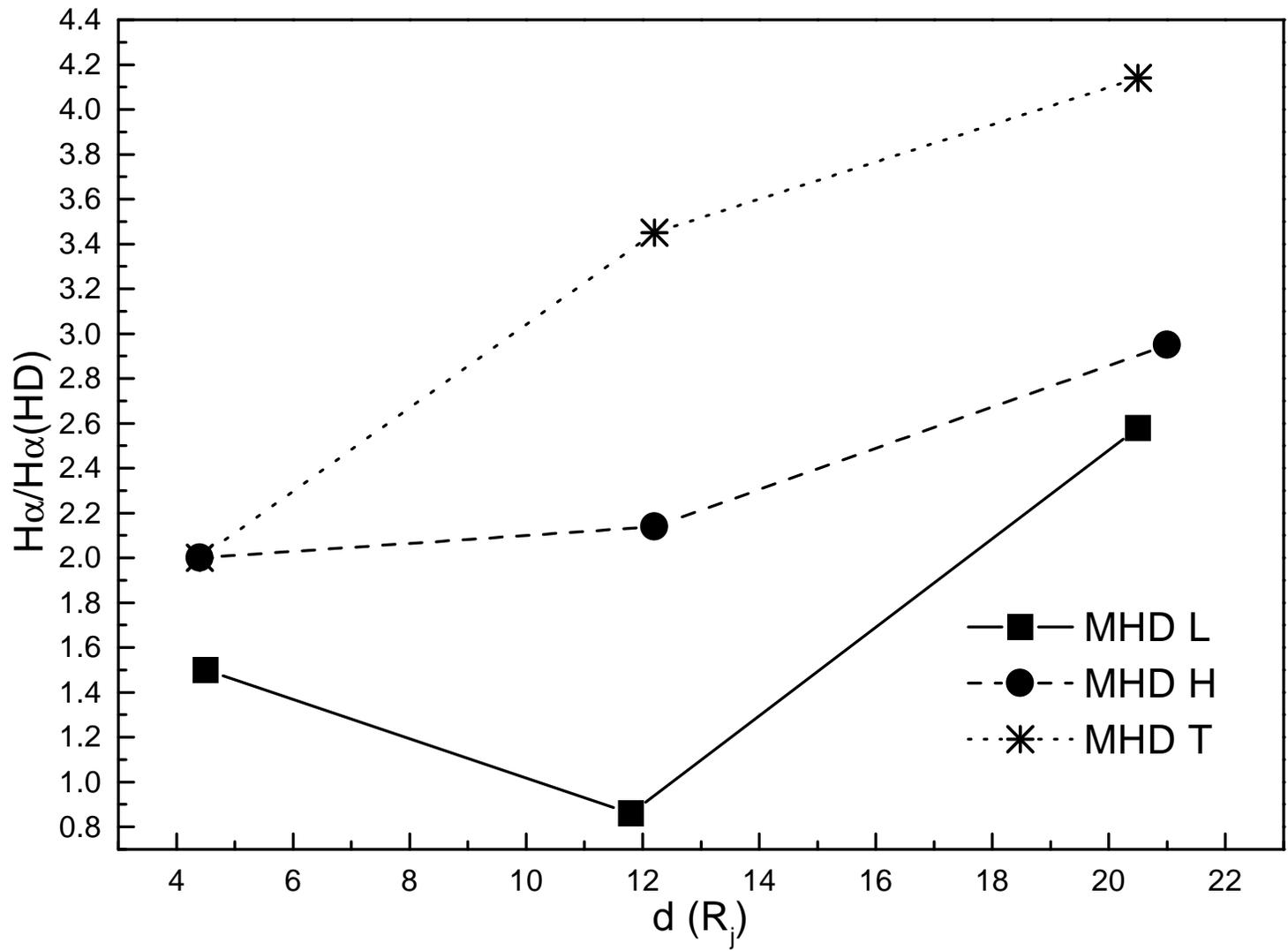
\end{document}